\documentclass[fleqn,usenatbib]{mnras}

\usepackage{newtxtext,newtxmath}

\usepackage[T1]{fontenc}
\usepackage{ae,aecompl}

\usepackage{graphicx}	
\usepackage{amsmath}	

\usepackage{amssymb}	
\usepackage{bm}    
\usepackage{url}
\usepackage{subfig}
\usepackage{longtable}
\usepackage{booktabs,caption}
\usepackage[flushleft]{threeparttable}
\usepackage{fixltx2e}
\usepackage{adjustbox}
\usepackage{multirow}
\usepackage{ltablex}

\title[The size-luminosity relation at z$=6-9$]{The size-luminosity relation of lensed galaxies at $z=6-9$ in the Hubble Frontier Fields }

\author[L. Yang et al.]{
Lilan Yang $^{1}$\thanks{JSPS Fellow; lilan.yang@ipmu.jp},
Nicha Leethochawalit $^{2,9,10}$
Tommaso Treu,$^{3}$,
Guido Roberts-Borsani $^{3}$,
\newauthor
Maru\v{s}a Brada\v{c} $^{4,11}$,
Simon Birrer $^{5,12}$,
Marco Castellano $^{6}$,
Emiliano Merlin $^{6}$,
\newauthor
Adriano Fontana $^{6}$,
Ricardo Amorin $^{7,8}$,
Michele Trenti $^{2}$
\\
\\
$^{1}$ Kavli Institute for the Physics and Mathematics of the Universe, The University of Tokyo, Kashiwa, Japan 277-8583 \\
$^{2}$ School of Physics, Tin Alley, University of Melbourne, VIC 3010, Australia\\
$^{3}$Department of Physics and Astronomy, University of California, Los Angeles, CA 90095-1547, USA\\
$^{4}$Department of Physics, University of California, Davis, CA 95616, USA\\
$^{5}$Kavli Institute for Particle Astrophysics and Cosmology and Department of Physics, Stanford University, Stanford, CA 94305, USA\\
$^{6}$INAF-Osservatorio Astronomico di Roma, Via Frascati 33, 00078 Monte Porzio Catone, RM, Italy\\
$^{7}$Departamento de Astronom\'ia, Universidad de La Serena, Av. Juan Cisternas 1200 Norte, La Serena, Chile\\
$^{8}$Instituto de Investigaci\'on Multidisciplinar en Ciencia y Tecnolog\'ia, Universidad de La Serena, Ra\'ul Bitr\'an 1305, La Serena, Chile\\
$^{9}$ ARC Centre of Excellence for All Sky Astrophysics in 3 Dimensions (ASTRO 3D), Australia\\
$^{10}$ National Astronomical Research Institute of Thailand (NARIT), MaeRim, Chiang Mai, 50180, Thailand\\
$^{11}$ University of Ljubljana, Department of Mathematics and Physics, Jadranska 19, 1000 Ljubljana, Slovenia\\
$^{12}$ SLAC National Accelerator Laboratory, Menlo Park, CA, 94025 
}

\date{Accepted XXX. Received YYY; in original form ZZZ}

\pubyear{2020}

\begin{document}
\label{firstpage}
\pagerange{\pageref{firstpage}--\pageref{lastpage}}
\maketitle
\begin{abstract}
We measure the size-luminosity relation of photometrically-selected galaxies within the redshift range $z\sim6-9$, using galaxies lensed by six foreground Hubble Frontier Fields 
(HFF) clusters. 
The power afforded by strong gravitational lensing allows us to observe fainter and smaller galaxies than in blank fields.  
We select our sample of galaxies and obtain their properties, e.g., redshift, magnitude, from the photometrically-derived ASTRODEEP catalogues. 
The intrinsic size is measured with the \texttt{Lenstruction} software,  and completeness maps are created as a function of size and luminosity via the \texttt{GLACiAR2} software. 
We perform a Bayesian analysis to estimate the intrinsic and incompleteness-corrected size-luminosity distribution, with parameterization $r_e \propto L^\beta$.  
We find slopes of $\beta\sim0.48\pm0.08$ at $z\sim6-7$ and $\beta\sim0.68\pm0.14$ at $z\sim8.5$, adopting the Bradac lens model.
The slope derived by lensed galaxies is steeper than that obtained in blank fields and is consistent with other independent determinations 
of the size-luminosity relation from the HFF dataset. 
We also investigate the systematic uncertainties correlated with the choice of lens models, 
finding that the slopes of size-luminosity relations derived from different models are consistent with each other, i.e. the modeling errors are not a significant source of uncertainty in the size-luminosity relation.

\end{abstract}




\section{Introduction}
Galaxy sizes are a fundamental observable for the study of their formation and evolution.
For disk galaxies it is generally believed that the size is driven by  the angular momentum acquired from their dark matter halos \citep{Fall1980, Mo1998, Dutton2007, Wechsler2018}.
For elliptical galaxies, other physical processes such as galaxy major and minor mergers, and  feedback from a central supermassive black hole or from supernovae \citep{Lopez2012, Choi2018, Scannapieco2008} are believed to play an important role.

The evolution of galaxy sizes over cosmic time and their correlation with other physical parameters are powerful tools to constrain the mechanism of galaxy growth, and has implications for our understanding of cosmic reionization \citep{Oesch2010}. 

On the one hand, the size evolution of different populations, e.g., early-type and late-type galaxies, represents a key benchmark for any model of galaxy formation and evolution. In a pioneering study
using data by the Sloan Digital Sky Survey, \citet{Shen2003} reported that the size of local ($z\sim0$) galaxies can be described by a log-normal distribution, where more massive galaxies display larger sizes and the size-mass relation of early-type galaxies is steeper than their late-type counterparts. At higher redshift (e.g., $z\sim2-3$), the size-mass relation remains consistent with the results measured in the local Universe, albeit with smaller sizes on average \citep{Trujillo2006, Mosleh2012, vanderwel2014, Yang2021}.
Such observations and results have been independently confirmed by numerous works \citep{Williams2010, Carollo2013,  Mosleh2017, Whitaker2017, Mowla2019, Kawinwanichakij2021}  
These kind of measurements have been used to test and rule out galaxy growth models, such as the simple monolithic collapse model for quiescent galaxies \citep{vanDokkum2008}, and specific size growth mechanisms, e.g., major or minor mergers \citep{Tacconi2008, Wellons2016}.
 
On the other hand, the size distribution of high redshift galaxies and their scaling relation with luminosity have important implications for our understanding of cosmic reionization.  
At $z\sim6$, the reionization process is largely considered to have finished according to multiple observation \citep[e.g.,][]{Fan2006,Mason2018}, but the details of the beginning and progress remain a puzzle.  
Observations with the Hubble Space Telescope (HST) have shown that the faint-end slope of the UV luminosity function (LF) is steep,
implying that faint galaxies likely provide the majority of ionizing photons that power the Universe's last phase transition.
For instance, using HST data \citet{Grazian2011, Grazian2012} pointed out the importance of the intrinsic galaxy size distribution for the determination of the UV LF faint-end slope. \citet{Grazian2011} showed the LF depends critically on the size used to simulate the observational completeness, i.e., the faint-end slope varies from -2.02 to -1.65 adopting more compact morphologies. The observed half-light radii of galaxies at $z\sim$7 are typically of order 0.5 kpc \citep{Shibuya2015, Paulino-Afonso2018}. 
Such compact galaxies are barely resolved even with the Hubble Space Telescope and the James Webb Space Telescope.

Magnification afforded by strong gravitational lensing allows us to have higher source sensitivity and better angular resolution for a given instrumental set-up.
The Hubble Frontier Fields (HFF) program \citep{Coe2015, Lotz2017} has delivered significant samples of extremely faint galaxies, pushing down to unprecedented depths and out to very high redshift. 
Taking advantage of HFF data, several studies have investigated the size-luminosity distribution of extremely faint galaxies, i.e., $M_\text{UV} > -16$, and its connection to the luminosity function \citep{Kawamata2015, Ishigaki2015, Laporte2016, Bouwens2017, Kawamata2018, Yue2018}. 
As a complimentary to HFF, 
\citet{Neufeld2021} investigated the size-magnitude measurement of galaxies in shallower but wider covering area of Reionisation Lensing Cluster Survey at z $>5.5$.
\citet{Kawamata2018} measured the size of lensed galaxies taking advantage of the complete HFF data sets (both six cluster centers and associated parallel fields), pushing the faint-end limit to absolute UV magnitudes ($M_\text{UV}$) of $\sim-12$, and finding a steep size-luminosity relation that implies a relatively shallow faint-end slope.
During the completion of this manuscript, a paper by \citet{Bouwens2021} that revisits the size-luminosity using the same HFF dataset, was published.

A crucial aspect of using strong lensing magnification to measure the size luminosity relation at high redshift is the robustness with respect to residual uncertainties in lens modeling.
For example, using simulated cluster lenses,
\citet{Meneghetti2017} investigated the accuracy and precision of lens models which based on different strong lensing methods,
i.e., parametric and non-parametric methods, and found a good agreement in general, even though the details of the magnification can change significantly especially in regions of high magnification. 
On the observational front, \citet{Yang2021} investigated the impact of the choice of lens models on derivations of the size-mass relation at redshifts of $z\sim1-3$, finding good agreement between results.
 \citet{Bouwens2021} minimizes the effects of systematic uncertainties in the lens model on the size-luminosity relation by taking the median magnification from all available parametric models.

Extending the work of \citet{Yang2021}, in this paper we revisit the size-luminosity relation using galaxies lensed by six HFF clusters at $z>6$, and
quantify how adopting different parametric and non-parametric lens models affects the results. 
We find a steep size luminosity relation consistent with that found by \citet{Kawamata2018} and \citet{Bouwens2021}. We show that the results are robust with respect to the choice of lens model, even though the prediction of each lens model for individual sources can be quite different. On aggregate the slope and intercept of the correlation vary by less than the statistical uncertainties.
 
The structure of this paper is as follows.
We describe the data and lens model in~Section~\ref{sec:data}. We 
measure the size of lensed galaxies in Section~\ref{sec:size-determ} and we 
derive the size-luminosity distribution at Section~\ref{sec:size-lum}. We  
compare results between lens models in Section~\ref{sec:mod-compare},
and present our discussion and conclusion at Section~\ref{sec:discussion} and Section~\ref{sec:conclusion}, respectively.

\section{Data}\label{sec:data}
In this section, we briefly introduce the data and lens models used in this work. 
We select our sample based on the source catalogue provided by the ASTRODEEP project.

\subsection{HFF data \& lens models}

The HFF program observed deep fields centered at six lensing clusters, Abell 2744, MACS J0416.1-2403, MACS J0717.5+3745, MACS J1149.5+2223, Abell S1063, and Abell 370, as well as their parallel fields.
The datasets include ACS (F435W, F606W, F814W) and WFC3/IR (F105W, F125W, F140W, F160W) imaging data.
We refer the reader to the HFF website \footnote{\url{https://outerspace.stsci.edu/display/HPR/HST+Frontier+Fields }} to obtain more general information about the project.
We adopt the Version 1.0 release of HFF mosaics images and the lens models provided by the community, which are public available \footnote{\url{https://archive.stsci.edu/prepds/frontier/}}, as an input to our analysis.
 
 The lens models of the six clusters are provided by five independent groups, henceforth, Bradac \citep{Bradac2005, Bradac2009, Hoag2016}, Williams \citep{Liesenborgs2006, Mohammed2014, Jauzac2014, Grillo2015}, CATS \citep{Jullo2009, Jauzac2012, Richard2014, Jauzac2014}, Zitrin \citep{Zitrin2009, Zitrin2013, Merten2009, Merten2011} and Sharon \citep{Jullo2007, Johnson2014}, respectively. 
 Those lens teams provide models of the cluster center adopting common inputs, although their modeling techniques (e.g., parametric and non-parametric methods) are distinct. 
One of the goals in this work is to investigate the systematic uncertainties from the selection of lens model, 
so, we focus on the galaxies in the cluster center that is covered by all five lens models.

\subsection{Sample selection}
We select our sample of galaxies at redshift $z>6$ based on the multi-wavelength photometric catalogues of these images from the ASTRODEEP project \footnote{\url{http://www.astrodeep.eu/}} \citep{Merlin2016, Castellano2016, Criscienzo2017, Shipley2018, Bradac2019}. 
The photometric catalogues are generated via combining all the available data including 
the data from HST, the Spitzer Space Telescope (Infrared Array Camera 3.6$\mu m$ and 4.5$\mu m$ imaging), and 
Deep $K_s$ images from Very Large Telescope (VLT) High-Acuity Wide-field K-band Imager (HAWK-I).
The catalogues inclue photometrically-derived global galaxy properties (e.g., redshift, stellar mass, star formation rate) from several studies derived with spectral energy distribution (SED)-fitting codes. A few galaxies in our sample have spectroscopic confirmed redshift \citep{Vanzella2021}.

Here, we briefly summarize how the catalogues were obtained, and 
we refer the reader to the ASTRODEEP papers for details.
First, they remove the intra-cluster light. 
Then, \textsc{SExtractor} software is used on the processed image with a \texttt{HOT+COLD} two detection models.
The ASTRODEEP project paid attention primarily to ther WFC H160 band to keep consistent with CANDELS and 3D-HST surveys \citep{Merlin2016}.
In this work, we perform size measurements based on processed H-band image to probe the rest-frame UV properties.
As complementary, to achieve deeper detection, the source detection is also performed in a stack of infrared images using the same strategies as single H-band.
The detection catalogue was then used to gain the photometric measurement via \textsc{SExtractor}, e.g., flux and its uncertainties, and evaluate high-level data products such as the photometric redshifts and rest-frame properties, 
and infrared photometry was measured using \textsc{T-PHOT} \citep{Merlin2015}.
We correct the flux in catalogue taking K-correction and lensing magnification derived from the lens model into consideration.

We select our sample from the parent ASTRODEEP catalogues with additional criteria of 1) having a reliable photometric redshift estimation, specifically RELFLAG=1 labeled in the catalogues,
2) no proximity to a particularly bright object, e.g., at least 1 arcsecond away from the brightest cluster galaxies, or the edge of the detectors.
In the end, our sample contains 140 galaxies at z $\sim6-7$, and 33 galaxies at z $\sim8.5$.

\section{Intrinsic size measurement}\label{sec:size-determ}
To obtain the intrinsic size of the lensed galaxies, we employ the  \texttt{python} softwares \texttt{Lenstruction}\footnote{\url{https://github.com/ylilan/lenstruction}}/ \texttt{Lenstronomy}\footnote{\url{https://github.com/sibirrer/lenstronomy} } which adopt a forward modeling technique to reconstruct the source brightness distribution correcting the lensing and blurring effects simultaneously. 
\texttt{Lenstruction} \citep{Yang2020} is a wrapper of the strong lensing software \texttt{Lenstronomy} \citep{Birrer2015, Birrer&Amara2018, Birrer2021}
that provides the interface to handle the cluster-scale source reconstruction and local lens model perturbation . 

The sizes derived by \texttt{Lenstruction/Lenstronomy} are robust \citep{Yang2021, Kawinwanichakij2021}, and \citet{Yang2021} demonstrated that the sizes measured by \texttt{Lenstruction/Lenstronomy}
are consistent with those measured by traditional softwares such as GALFIT \citep{Peng2002}.

\subsection{Configuration of the data, source and lens models }

The inputs to the forward modeling  consist of a configuration of observed data, e.g., frame size of data, exposure time,  selection of a galaxy light profile, and lens model setup.
We measure the size of each galaxy in a $2.1"\times2.1"$ stamp from H-band processed image, i.e., intra-cluster light subtracted image,  that is used to perform source detection by ASTRODEEP. 
In some cases, the field of the stamp contains foreground contamination such as lens plane light.
\texttt{Lenstruction} can mask out the contamination region or fit them by additional light profiles.

We adopt the standard elliptical S\'ersic light profile to describe the background galaxies in the source plane. Its free parameters consist of the surface brightness amplitude, 
the circularized half-light radius r$_\text{e}$,  r$_\text{e}$ =  r$_\text{e}^\text{maj}\sqrt{1-e}$ where $e$ is the ellipticity, and r$_\text{e}^\text{maj}$ is the radius along the major axis,
 Se\'rsic index (n$_\text{s\'ersic}$), axial ratio (q), position angle, 
and the position of the center of light.

For every lens model, we simulate the lensed images and convolve them with the point spread function (PSF) in the image plane. We select a bright star in the same field of the targeted galaxy as the PSF.
The posterior distribution of the parameters are obtained by comparing the simulated and observed image, and considering the noise in each pixel with a combination of Gaussian background r.m.s. fluctuations and Poisson noise from the source counts.

In this exercise, since galaxies are small and generally have insufficient signal to noise ratio to determine the S\'ersic index,  we fix it to 1, appropriate for star-forming galaxies. The axis ratio is limited to be in the range 0.1-1.

We use lens models from those five modeling teams, i.e., Bradac, Williams, CATS, Zitrin and Sharon, respectively.
We assume that lens model parameters are smooth over the area spanned by each observed image. 
Thus, lens mode parameters are taken directly from the maps provided by above teams.

\subsection{Modeling procedures}
To constraint the source light profile, \texttt{Lenstrcution} adopts 
Bayesian inference to estimate the posterior distribution of the free parameters of 
the source model.
For a given lens model, an initial fit of the source parameters is found first via linear minimization,
i.e, comparing the modeled image with observation. Once the initial fit is found, we run Markov chain Monte Carlo (MCMC) process to  explore the full posterior and provide confidence intervals of source parameters \citep[\texttt{emcee},][]{emcee}. 

To explore the discrepancy between lens models, we repeat a similar modeling process for each lensed galaxy,  while changing the input lens model parameters. The details of comparison are presented in Section~\ref{sec:mod-compare}. The results are shown in Table~\ref{tab:fits-z67} and Table~\ref{tab:fits-z8} for two redshift bins.

\begin{figure*}  
\centering
\includegraphics[width=1.8\columnwidth]{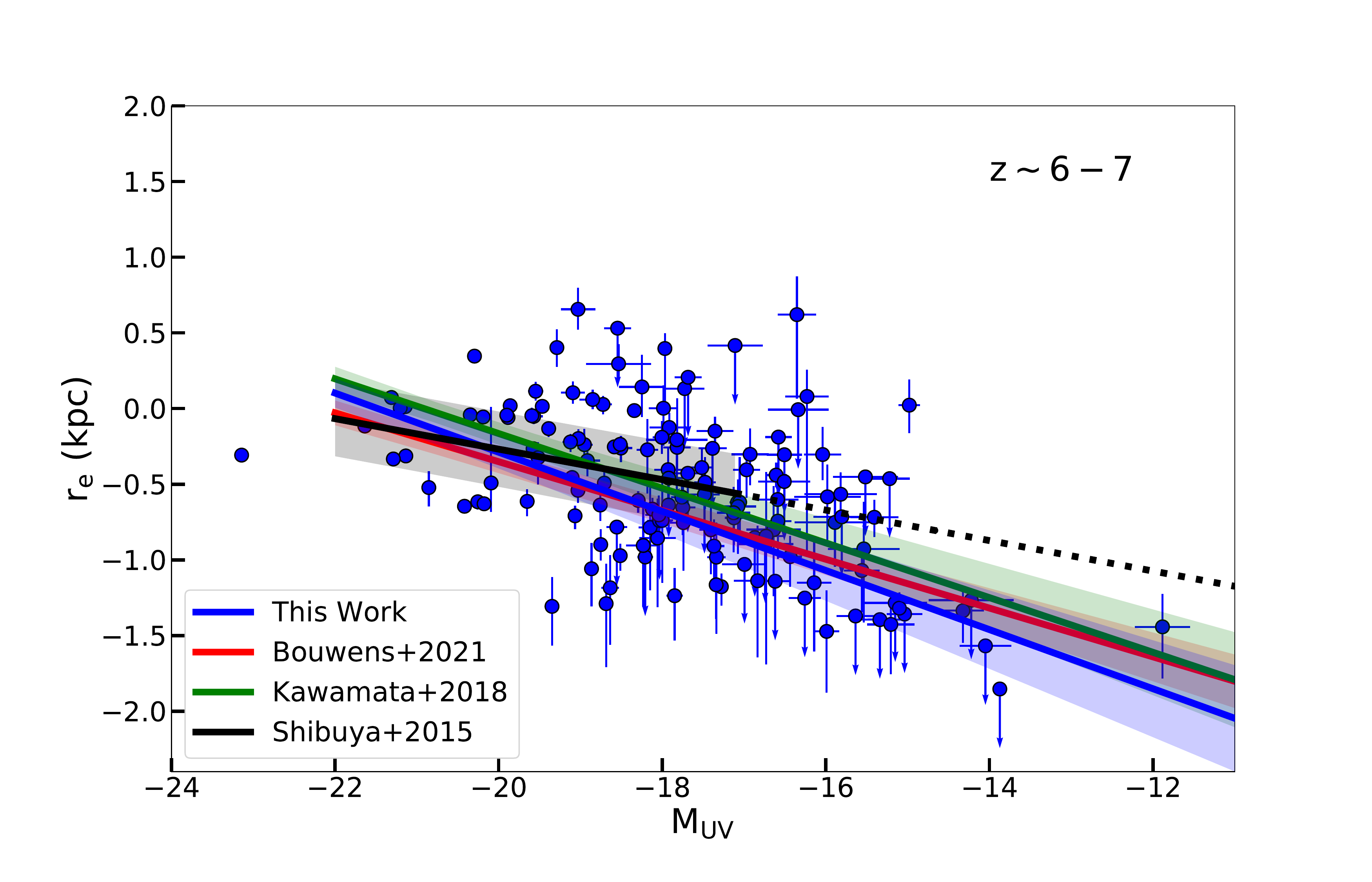} 
\includegraphics[width=1.8\columnwidth]{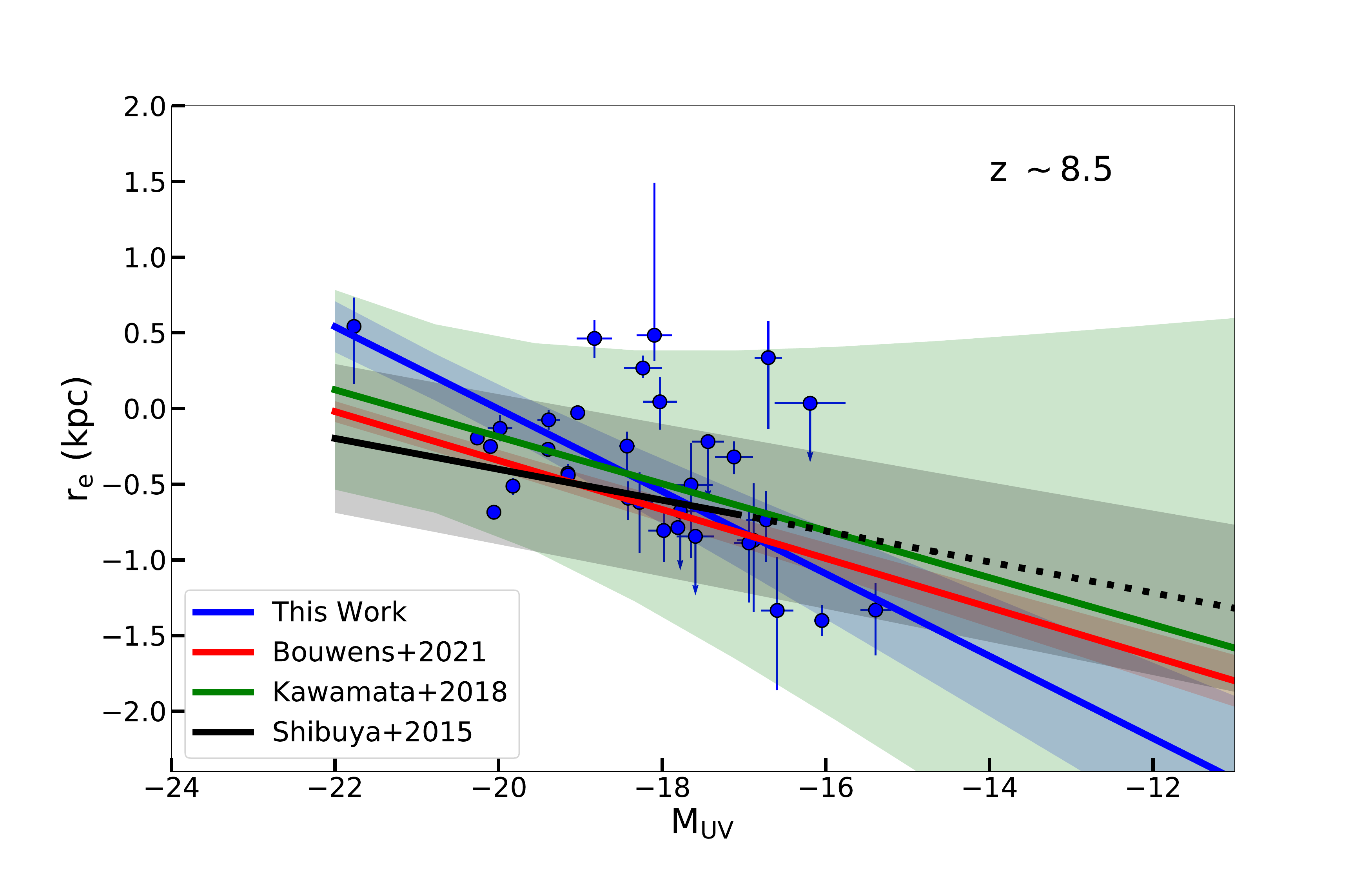} 
\caption{Intrinsic size-luminosity distribution of galaxies lensed by the six HFF clusters at z $\sim6-7$ (top) and z$\sim$8 (bottom). The blue points represent measurements from this work, see details in Table~\ref{tab:fits-z67} and Table~\ref{tab:fits-z8}, adopting the Bradac lens model. 
The solid lines in each panel represent the best-fits of the  size-luminosity relation obtained by this work (blue), two other consistent results derived from lensed galaxies \citet{Bouwens2021} (red) and \citet{Kawamata2018} (green), and result derived from the blank field \citet{Shibuya2015} (black), respectively. The shadow regions show the 1$\sigma$ distribution of those. The fitting results are shown in Table~\ref{tab:results}.  }

\label{fig:size-muv}
\end{figure*}

\begin{table*}
\centering
\caption{Best-fits parameters of size-luminosity distribution at z $\sim6-7$ and z $\sim8.5$ for five lens models.}

\begin{tabular}{ l c c c }
\hline
\hline
Lens model & lg$r_0$ & $\sigma$ & $\beta$ \\
\hline
z $\sim6-7$\\
Bradac &-0.16 $\pm$ 0.11 & 1.33 $\pm$ 0.08 & 0.48 $\pm$ 0.08 \\
Williams & -0.13 $\pm$ 0.1 & 1.34 $\pm$ 0.08 & 0.49 $\pm$ 0.08   \\
CATs &0.00 $\pm$ 0.08 & 1.21 $\pm$ 0.07 & 0.62 $\pm$ 0.08  \\
Zitrin  & -0.05 $\pm$ 0.09 &1.03 $\pm$ 0.06 & 0.47 $\pm$ 0.06  \\ 
Sharon & -0.07 $\pm$ 0.1 & 1.18 $\pm$ 0.07 & 0.50 $\pm$ 0.07  \\

 \hline
z $\sim8.5$\\
Bradac & 0.21 $\pm$ 0.17 & 1.02 $\pm$ 0.13 & 0.68 $\pm$ 0.14 \\  
Williams &0.16 $\pm$ 0.18 & 1.44 $\pm$ 0.19 & 0.73 $\pm$ 0.16 \\
CATs & 0.12 $\pm$ 0.13 & 1.21 $\pm$ 0.16 & 0.78 $\pm$ 0.15 \\
Zitrin & 0.17 $\pm$ 0.19 & 1.27 $\pm$ 0.17 & 0.64 $\pm$ 0.13 \\    
Sharon &0.12 $\pm$ 0.17 & 1.06 $\pm$ 0.14 & 0.58 $\pm$ 0.14  \\
  \hline
\end{tabular}

\label{tab:results}
\end{table*}

\section{Size-Luminosity distribution}\label{sec:size-lum}
The observed size-luminosity distributions of galaxies lensed by HFF clusters at z $\sim6-7$ and z $\sim8.5$ are  shown in  Figure~\ref{fig:size-muv}.
The x-axis represents the corrected UV magnitude from the ASTRODEEP catalogues, 
and the y-axis shows the intrinsic size measured by \texttt{Lenstruction}.
Visually, we can tell that the more luminous galaxies tend to have a larger size.
Lensing provides the ability to observe very faint and small galaxies.
The faint magnitude limit reaches $\sim$-13 and the resolvable size reaches tens of pc, similar to the limits limits reported by \citet{Kawamata2018}.

We present our method to derive the intrinsic size-luminosity relation in Section ~\ref{sec:analy},
and we perform the incompleteness correction based on the \texttt{python} code \texttt{GLACiAR2} \citep{Carrasco2018,Leethochawalit2021} in Section ~\ref{sec:comp-corr}.
Finally, in Section ~\ref{sec:intrinsic-sl}, we present the results of the lensing and incompleteness corrected size-luminosity relation.

\subsection{Analytical description} \label{sec:analy}

The probability density function (PDF) of the size-luminosity pair is described as (see details in \citet{Kawamata2018}), 
\begin{equation}\label{eq:lognormal}
\centering
\begin{split}
\Psi_\text{obs}(r_{e}, M_\text{UV})&=\Psi_\text{ins}(r_{e}, M_\text{UV}; r_{0}, \sigma, \beta, M^{*}, \alpha)C(r_{e}, M_\text{UV}) \\
&=P(r_{e}, M_\text{UV}; r_{0}, \sigma, \beta)\phi(M_\text{UV}; M^{*}, \alpha)C(r_{e}, M_\text{UV}) \\
\end{split}
\end{equation}
where $\Psi_\text{obs}$ and $\Psi_\text{ins}$ are the observed and intrinsic PDF, respectively,
and $C(r_{e}, M_\text{UV})$ is the completeness map as a function of size and UV magnitude, $M_\text{UV}$, for each cluster, see details in Section~\ref{sec:comp-corr}.
$\Psi_\text{ins}$ is determined jointly by the PDF of the size distribution $P(r_{e}, M_\text{UV}; r_{0}, \sigma, \beta)$ and that of the luminosity  $\phi(M_\text{UV}; M^{*}, \alpha)$.
$P(r_{e}, M_\text{UV}; r_{0}, \sigma, \beta)$ is described by a log-normal distribution
\begin{equation*}
 P(r_{e}, M_\text{UV}; r_{0}, \sigma, \beta)=\frac{1}{r_e\sigma\sqrt{2\pi}} exp \left(-\frac{({\rm ln}r_e-{\rm ln}\overline{r_e})^2}{2\sigma^2}\right)
\end{equation*}
and
\begin{equation*}
\overline{r_e}=r_0\left (\frac{L}{L_0} \right)^\beta
\end{equation*}
where the r$_0$, $\sigma$, $\beta$ and $L_0$ are the radius at M$_\text{UV}$=-21, the 
variance of the log-normal distribution, the slope of the size-luminosity relation, and luminosity corresponding to M$_\text{UV}$=-21, respectively.

The luminosity function is described by the Schecheter function.
Since our sample is not large enough to determine the luminosity function as well as the size luminosity relation,  we fix the parameters  $M^*$ and $\alpha$ to -20.73 and -1.86 following \citet{Kawamata2018}.
They demonstrated that there is no significant difference in results between free or fixed parameters of the luminosity function.  

\subsection{Incompleteness correction}\label{sec:comp-corr}

In order to obtain the intrinsic size luminosity relation we have to account for the dependency of completeness on size and luminosity.
The completeness correction is computed using the software \texttt{GLACiAR2} \footnote{https://github.com/nleethochawalit/GLACiAR2-master}
\citep{Carrasco2018,Leethochawalit2021}, which is a public \texttt{python} tool for simulations of source recovery and completeness in galaxy surveys. 
The original code  \texttt{GLACiAR2} did not include the effects of gravitational lensing, therefore we modified it in order to apply it to the lensing fields.

The modified code performs the following steps to obtain the 2D completeness map $C(r_{e}, M_\text{UV})$, where the main concept is to calculate the fraction 
of simulated lensed galaxies that are detected and pass the selection criteria of ASTRODEEP catalogues.

1. We select random positions uniformly on the source plane. The allowable position area in the source plane is reconstructed from  mapping the image plane back to the source plane with the deflection map provided by each lens model team.

2. At each randomly selected position, we generate artificial galaxies with a range of sizes and magnitude. We set 7 bins in radius within the range $r_e\sim0.1-10$ kpc, and 7 magnitude bins $M_\text{UV}$ from -22 to -16.
Taking the lensing and PSF effects into account, we simulate how the galaxies would then appear in the F160W selection image used for catalog construction. 
The artificial galaxies are modeled with a S\'ersic profile, with the amplitude, effective radius $r_e$ given by input, 
and index $n_\text{S\'ersic}$ range 1-4.

3. The code runs \textsc{SExtractor} on the simulated image and calculates the fraction of 
artificial galaxies that are detected using same configuration as ASTRODEEP, 
i.e, \texttt{HOT+COLD} mode.

4. We loop through the input size and luminosity bins to derive the completeness map as a function of input size and UV magnitude.

5. We run the modified \texttt{GLACiAR2} for each cluster field adopting all the available lens models at $z\sim6-7$ and $z\sim8.5$.

As an example, the completeness maps for the cluster Abell 2744 for five lens models and $z\sim6-7$ are shown
in Figure~\ref{fig:a2744-comp-compare}.
As expected completeness decreases with decreasing luminosity and increasing size.

\begin{figure*} 
\centering
    \includegraphics[width=\columnwidth]{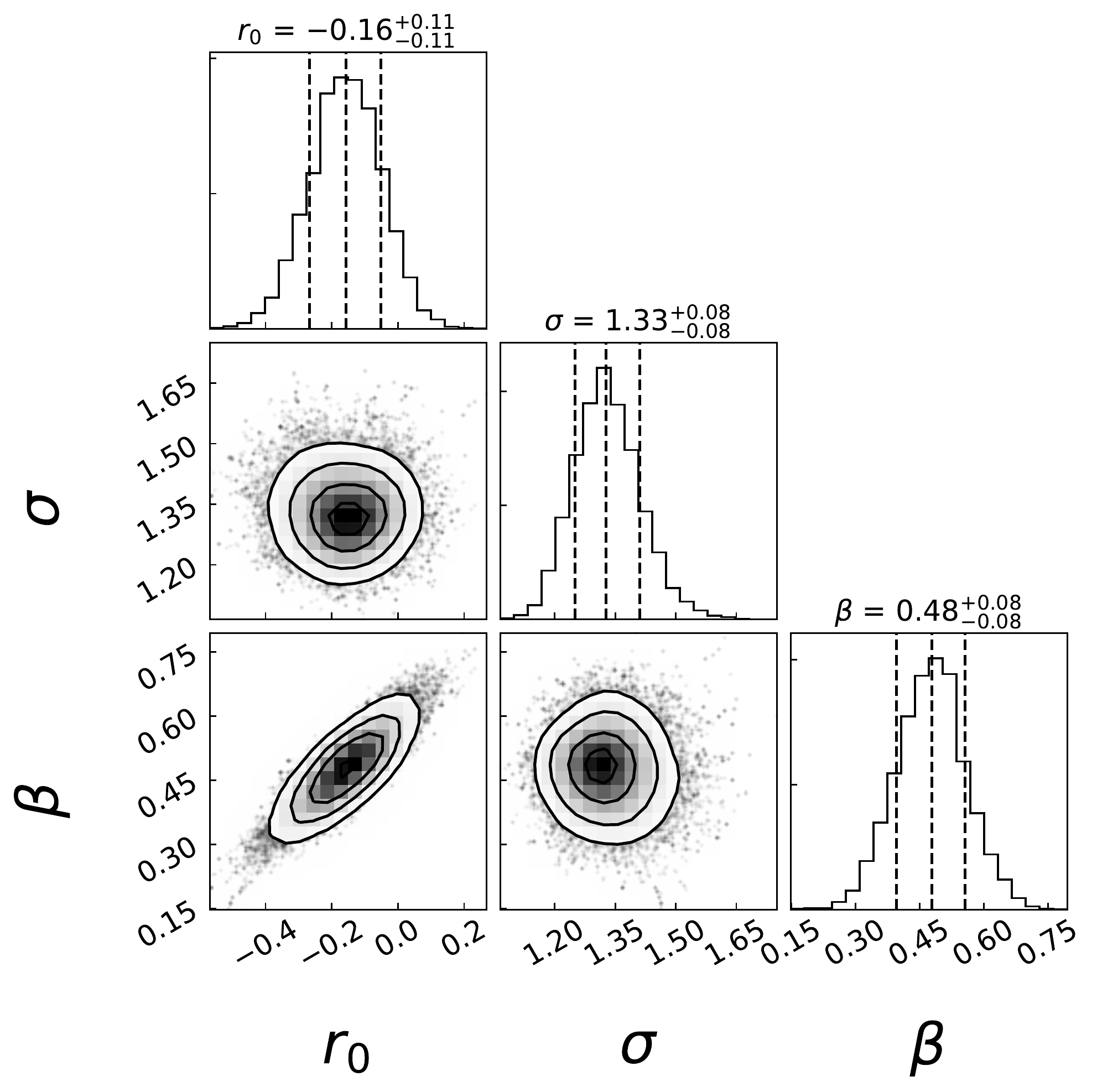}
    \includegraphics[width=\columnwidth]{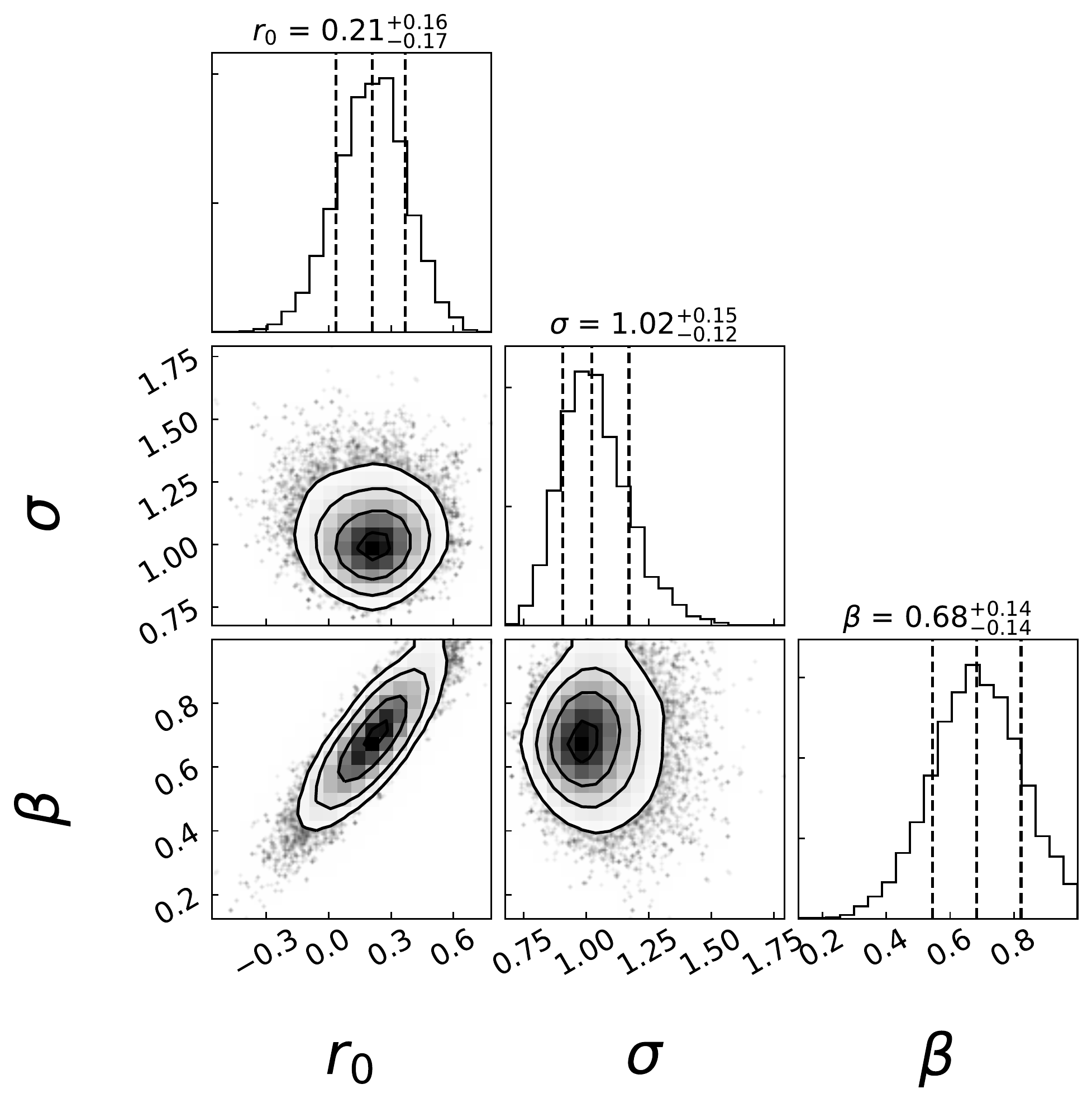}
    \caption{Posterior distribution of the r$_0$, $\sigma$ and $\beta$ parameters of the size-luminosity distribution at $z\sim6-7$ (left) and $z\sim8.5$ (right), adopting the Bradac lens model. The vertical lines in each panel display the best-fit value and 1$\sigma$ error.  }
    \label{fig:mcmc}
\end{figure*}

\subsection{Intrinsic size-luminosity relation}\label{sec:intrinsic-sl}
We use a standard Bayesian approach to derive the posterior distribution of the parameters describing the 
size-luminosity relation, r$_0$, $\sigma$, $\beta$, see Equation~\ref{eq:lognormal}. We assume a flat prior for those three parameters.
The best-fit results are shown in Figure~\ref{fig:size-muv}, 
and the MCMC results of those three parameters at $z\sim6-7$ and  $z\sim8.5$ are shown in Figure~\ref{fig:mcmc}. Details are given in Table~\ref{tab:results}.

In Figure~\ref{fig:size-muv} the solid lines and shaded region represent the best-fits and 1$\sigma$ error result of size-luminosity relation at
$z\sim6-7$ (left panel) and $z\sim8.5$ (right panel), adopting the Bradac lens model as an example. At  $z\sim6-7$, the best-fit slope $\beta$ of $r_e\propto L^\beta$ is $0.48\pm0.08$ for the Bradac model. 
We derive the size-luminosity relation using all five lens models, 
and the results are robust with respect to the choice of lens model as detailed in Section~\ref{sec:mod-compare}. 
The size at $M_\text{UV}=-21$ is $0.69\pm0.18$ kpc, and 
the variance of log-normal distribution $\sigma$ is $\sim1.33\pm0.08$.

At  $z\sim8.5$, the best-fit slope $\beta\sim0.68\pm0.14$ for the Bradac model, which is consistent with that found at $z\sim6-7$.
We also find excellent agreement between lens models as demonstrated in Figure~\ref{fig:sl-comparsion}.   
The 1$\sigma$ error is quite large, limited by sample size.
We only have a handful of galaxies with reliable size measurement in this redshift range, especially at the faint end, i.e., $M_\text{UV}>-16$.
The modal size is  $1.62\pm0.65$ kpc.
The best-fit values of the intrinsic scatter $\sigma$ in two redshift ranges are similar, and also
in good agreement with values determined by previous studies \citep{Kawamata2018}.

\section{Comparison between lens models}\label{sec:mod-compare}
\begin{figure*} 
\centering
    \includegraphics[width=1.8\columnwidth]{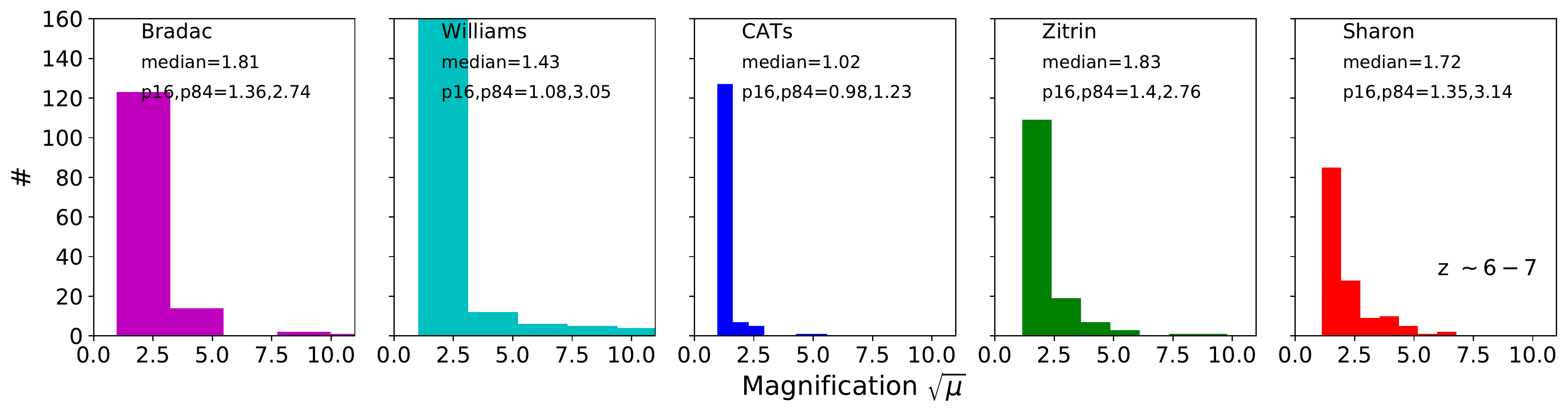} \\
     \includegraphics[width=1.8\columnwidth]{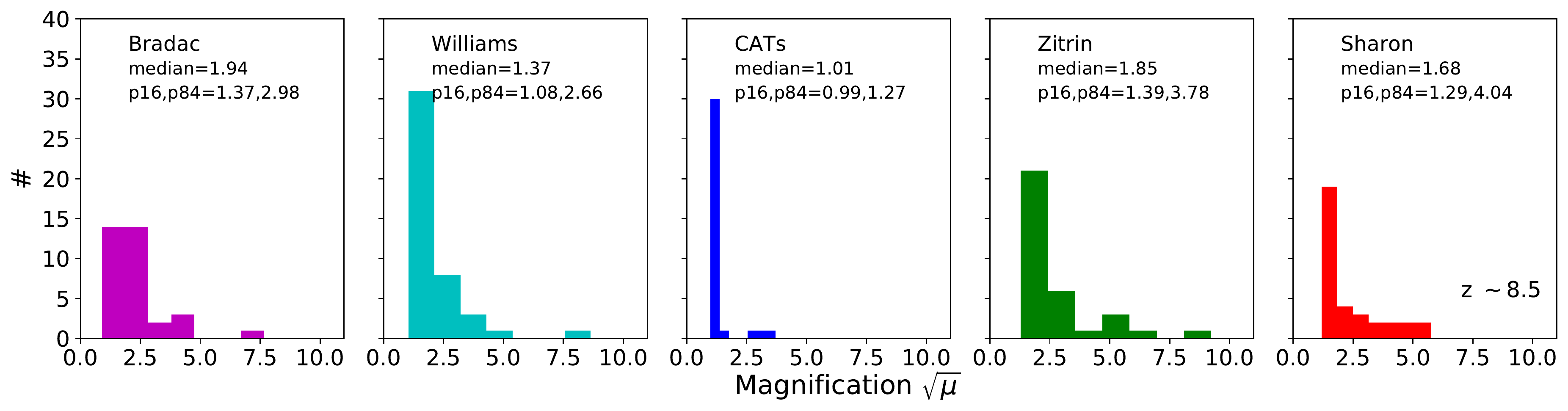}

    \caption{Comparison of the square root of magnification $\sqrt{\mu}$ between the five lens models at $z\sim6-7$ (upper) and $z\sim8.5$ (bottom).
    In each panel, we indicate the lens model, median value, and the 16th and 84th percentiles of the distribution.}
    
    \label{fig:all-magnif-compare}
\end{figure*}

\begin{figure*} 
\centering
\includegraphics[width=1.8\columnwidth]{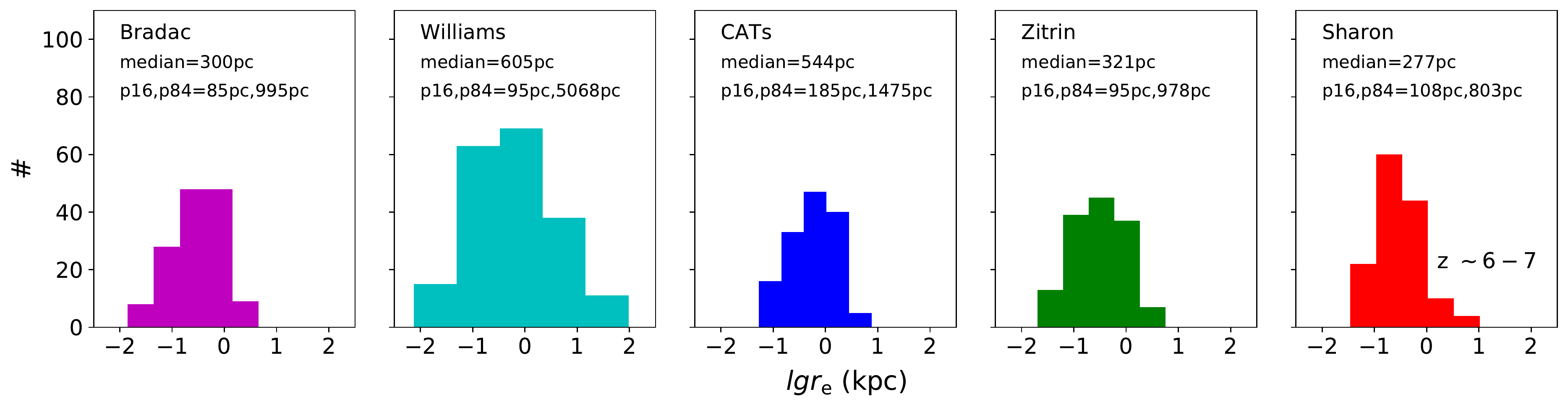}\\
    \includegraphics[width=1.8\columnwidth]{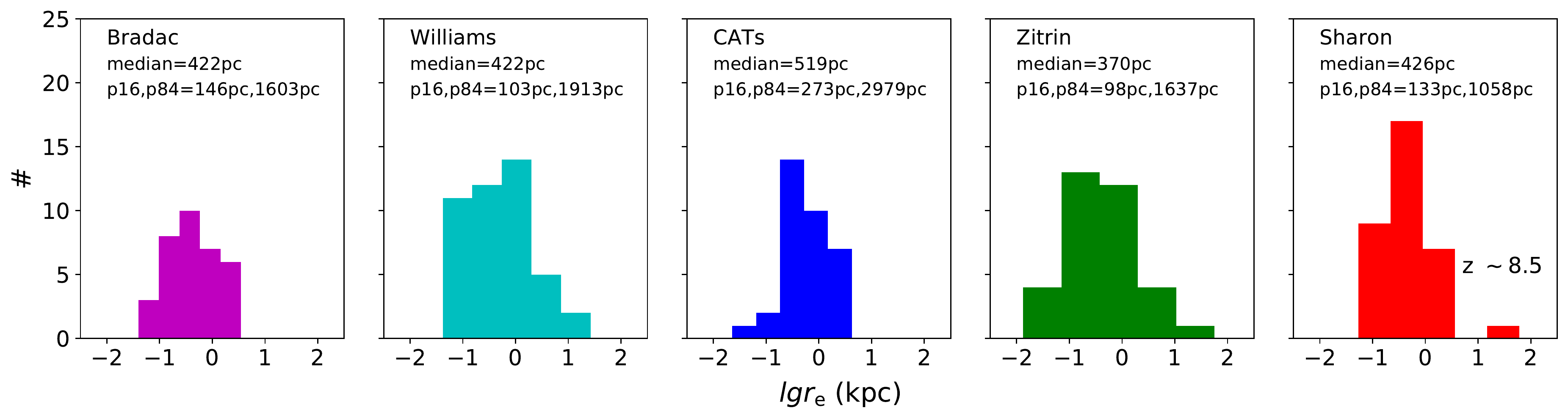}
        \caption{Comparison of reconstructed sizes $r_\text{e}$ between five lens models at $z\sim6-7$ (upper) and $z\sim8.5$ (bottom). 
       In each panel, we indicate the lens model, median value, and the 16th and 84th percentiles of the distribution.}
    \label{fig:all-size-compare}
\end{figure*}

\begin{figure*} 
\centering
    \includegraphics[width=1.8\columnwidth]{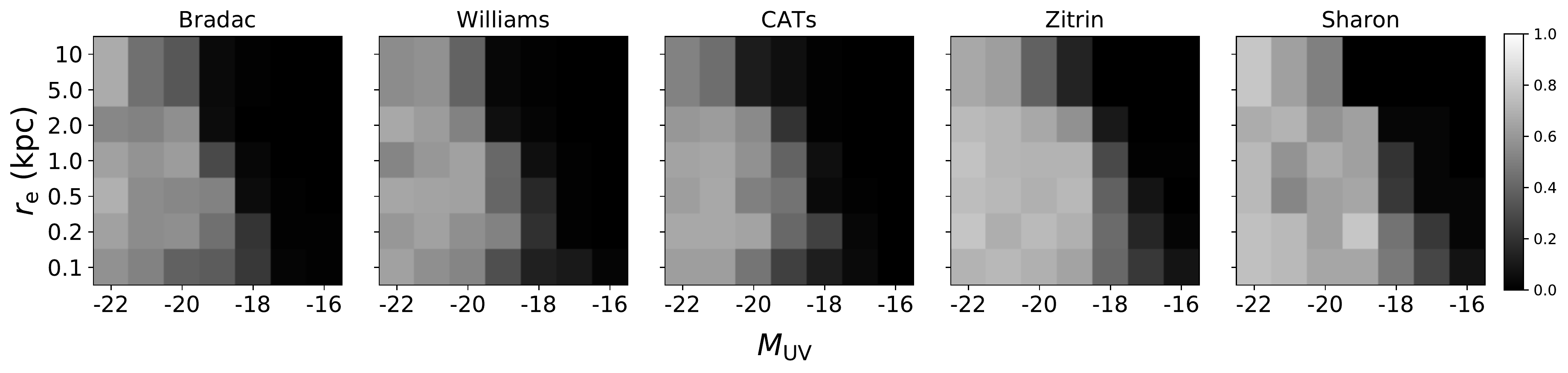}
    \caption{Completeness map of cluster A2744 for the five lens models at $z\sim6-7$. 
    The detection fraction is a function of UV magnitude and size. The color scheme represents the recovery fraction, darker colors corresponding to lower recovery fraction.}
    \label{fig:a2744-comp-compare}
\end{figure*}

\begin{figure*} 
\centering
    \includegraphics[width=\columnwidth]{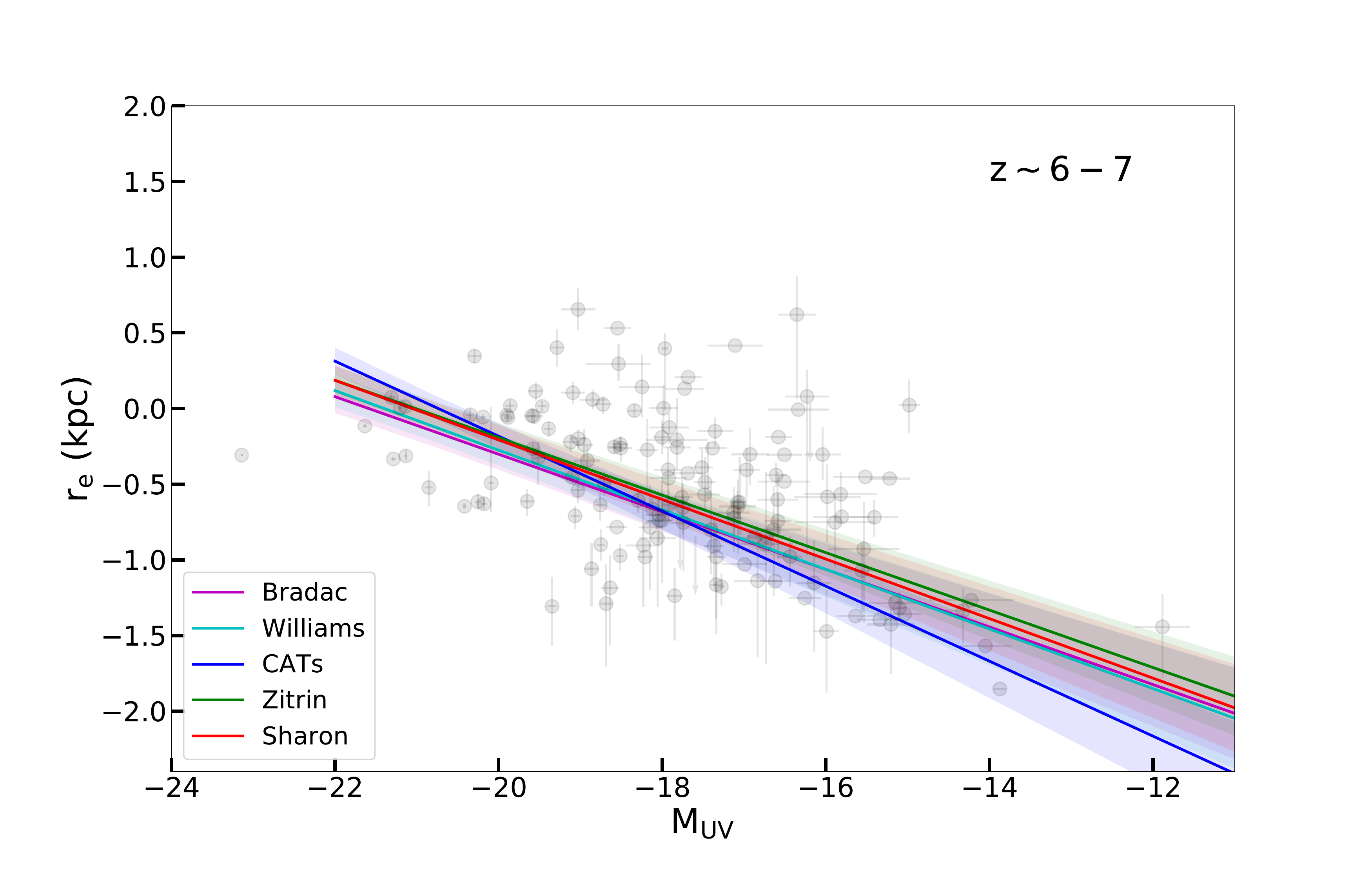}\\
    \includegraphics[width=\columnwidth]{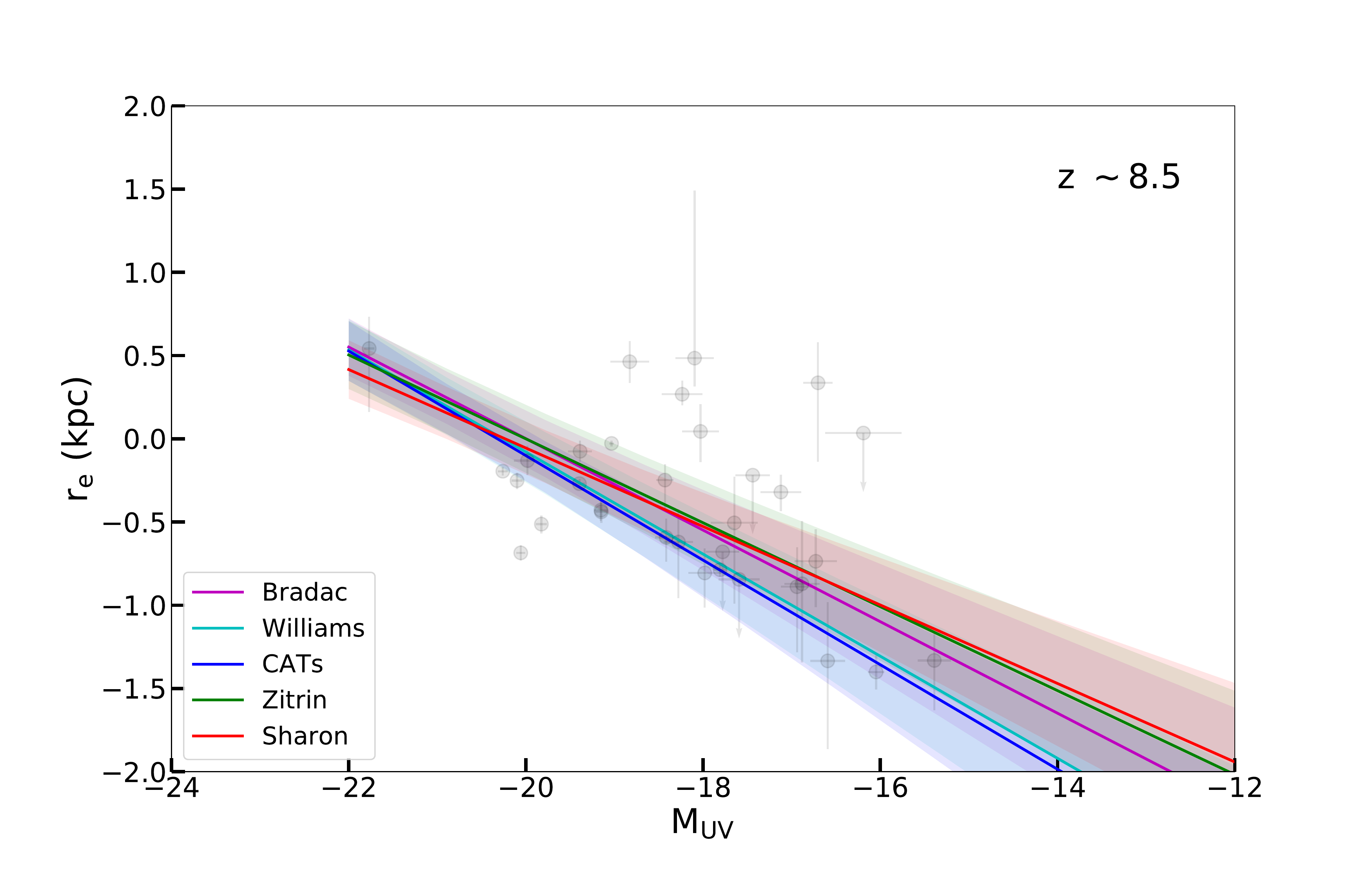}
    \caption{Comparison of the size-luminosity relations based on the different lens models. 
    The background grey points are the measurements from the Bradac lens model, same as Figure~\ref{fig:size-muv}. The solid lines and shaded areas indicate the best-fit and 1$\sigma$ error of size-luminosity relation. The colors represent different lens models as labeled. 
    }
    \label{fig:sl-comparsion}
\end{figure*}

In this section, we compare the results derived from different lens models, in terms of magnification, size, and completeness map as shown in 
Figure \ref{fig:all-magnif-compare}, Figure~\ref{fig:all-size-compare} and Figure~\ref{fig:a2744-comp-compare}, respectively.
Taking all factors into consideration, the comparison of the best-fits results of the size-luminosity distribution is presented in Figure~\ref{fig:sl-comparsion}.  

The magnifications derived from five lens models are shown in Figure~\ref{fig:all-magnif-compare}. 
We show the distribution of the square root of the magnification $\sqrt{\mu}$, which is proportional to the source stretch scale.  
We characterize the distributions of the $\sqrt{\mu}$ by their median and of their 16th and 84th percentiles, p16 and p84. 
At $z\sim6-7$ The mean $\sqrt{\mu}$ values of 3/5 models peak around $\sim1.8$, while 2/5 models peak  $\sim1$.
The behavior is similar at $z\sim8.5$.

The reconstructed size distributions for each of the five lens models are shown in 
panels of Figure~\ref{fig:all-size-compare}, and we characterize the distributions in the same manner as Figure~\ref{fig:all-magnif-compare}.
At $z\sim6-7$, the median $r_\text{e}$ values is around $\sim300-600$ pc.
The median value of  CATs is relatively larger, 543 pc, and
the p16 value of  CATs is 185 pc almost two times of other lens models,
which are the results of the small median $\sqrt{\mu}$.
We note that although Williams's model has a similar median value of  $\sqrt{\mu}$ and  $r_\text{e}$ as CATs, the p16 value is at the same level as the other three lens models, bacause the size stretch factor is also affected by the shear parameters \citep{Bouwens2021}.

The completeness map is computed for each cluster using each of the five lens models. As an example, cluster A2744 is shown in Figure~\ref{fig:a2744-comp-compare}. 
After applying the completeness correction, we compare the inferred size-luminosity relation in Figure~\ref{fig:sl-comparsion}.
The solids lines and shaded regions represent the best-fits and 1$\sigma$ error from each of the five lens models. The numerical results from the fits are listed in Table\ref{tab:results}.
The background grey points are reconstructed from the Bradac model as a reference.
As shown in Table\ref{tab:results}, the slopes based on the five lens modes are consistent with each other. We conclude that although there are discrepancies for individual sources, 
the inference of the size-luminosity relation is robust.

\section{Discussion}\label{sec:discussion}

In this section, we first compare our results with previous work,
and we then discuss the implications for the luminosity function.

\subsection{Comparison with previous works}

We have shown that the inferred size luminosity relation is robust with respect to the choice of lens model. 
We now compare our results with those obtained by other groups in blank and lensed fields.
In Figure~\ref{fig:size-muv}, the solid lines represent the results obtained by us (blue), by \citet{Bouwens2021} (red), \citet{Kawamata2018} (green) and   \citet{Shibuya2015} (black), respectively.
In blank fields, 
\citet{Shibuya2015} investigated $\sim190, 000$ HST galaxies with M$_\text{UV}<-18$ and
found a shallow slope of the size-luminosity relation with a constant slope $\beta=0.27\pm0.01$ at $0<z<8$.
This result is consistent with those of other studies in blank fields \citep[e.g.,][]{Huang2013}.

In contrast, in this work, we find a steeper slope $\beta\sim0.48\pm0.08$ and $\beta\sim0.68\pm0.14$ at  $z\sim6-7$ and $z\sim8.5$ (with Bradac lens model). These finding are  consistent with those of previous studies in lensed fields. 
\citet{Kawamata2018} used lensed galaxies in both the central and parallel fields of the six HFF clusters. 
They find that the slope at $z\sim6-7$ is steep, i.e., $\beta\sim0.46\pm0.09$, confirming their previous result based on a smaller sample 
 size \citet{Kawamata2015}.  
The slope does not vary significantly in the redshift range $z\sim6-9$, although the 1$\sigma$ errors at z$>7$ are larger due to the smaller sample size.
Very recently, an independent study \citep{Bouwens2021}  revisited 
the relation using a similar dataset and confirmed the steepness of the slope $\beta$.
They find $\beta\sim0.40\pm0.04$ at $z\sim6-8$.

The difference between blank and lens fields is due to a combination of two factors. First,  the lensed fields reach intrinsically fainter galaxies. Second, it is likely that the smallest galaxies are missed, or their sizes overestimated, in the blank fields.

\subsection{Implication for the faint-end slope of the luminosity function}

The intrinsic distribution of galaxy sizes is a crucial ingredient to determine the faint-end slope of the luminosity function, because of the effect of size on incompleteness correction, see Figure~\ref{fig:a2744-comp-compare}. In general, the smaller the galaxies, the smaller will be the completeness correction at the faint end and thus the shallower the inferred faint-end slope of the luminosity function. In turn, a shallower faint end slope will result in fewer UV photons down to a given absolute magnitude and thus require a higher escape fraction from a fainter minimum luminosity of galaxies to ionize the intergalactic medium. Of course, there are other factors to be taken into account. For example, photons might escape more easily from a very compact source, thus somewhat compensating the smaller numbers via a higher escape fraction.

In conclusion, our work confirms the steep slope of the size luminosity relation found by  \citet{Bouwens2021} and \citet{Kawamata2018}, demonstrating that the result is robust to the choice of lens model and it is thus not driven by lens mode uncertainties.

\section{Conclusion}\label{sec:conclusion}

In this paper, we revisited the size-luminosity relation of galaxies at the epoch of reionization. We focused on the six HFF cluster fields and 
compared the performance of five publicly available lens models. 
Our sample is selected from the multi-wavelength photometric catalogue made by the ASTRODEEP group, size measurements are 
conducted via the modified \texttt{Lenstruction} software and we correct for observational incompleteness via the \texttt{GLACiAR2} software.
We summarize our results as follows,

1. Parametrizing the size-luminosity relation as $r_e \propto L^\beta$, we find a slope  $\beta\sim0.48\pm0.08$ at $z\sim6-7$ and  
$\beta\sim0.68\pm0.14$  at $z\sim8.5$, adopting the Bradac lens model. The slope is steeper than that found in blank fields and consistent with that found by previous studies in lensing fields. 

2. The slope $\beta$ is robust with respect to the choice of lens model and varies by less than the reported uncertainties.

Our study demonstrates that the steeper slope found in lensing fields with respect to blank fields cannot be explained with lens modeling systematics. We attribute the steeper slope to the fainter luminosity and higher angular resolution achieved in lensing fields, which allows us to measure intrinsically smaller galaxy sizes at the faint end. 

The steep slope  of the size luminosity relation has implications for cosmic reionization as incompleteness depends on intrinsic galaxy size. Therefore, qualitatively, a steep slope of the size luminosity relation will result in a shallower faint end slope of the galaxy luminosity function. Studies with larger samples of lensed galaxies are needed to fully quantify this effect. 

The successful launch of the James Webb Space Telescope and the data taken under programs such as ERS-1324 will allow us to carry out such studies, by probing even fainter magnitudes and smaller sizes, and simultaneously constrain the size luminosity relation and the galaxy luminosity function.


\begin{table*}
\caption{Fitting results for galaxies lensed by six HFF clusters at  $z\sim6-7$, adopting the Bradac lens model.
From left to right, the columns represent the ID in the ASTRODEEP catalogue of source, RA and Dec of the center of light, photometric redshift, circularized effective radius $r_e$, and  UV magnitude. 
}
\setcounter{table}{1}
\label{tab:fits-z67}
\begin{tabular}{cccccc}
\hline
ID & RA & Dec & z$_\text{photo} $ & r$_e$/kpc & M$_\text{UV}$ \\
\hline
Abell 2744\\
62 & 3.604872 & -30.415848 & 6.13 & 0.91 $\pm$ 0.08 & -20.35 $\pm$ 0.04\\
73 & 3.593802 & -30.415447 & 6.85 & 0.13 $\pm$ 0.03 & -18.75 $\pm$ 0.05\\
106 & 3.570654 & -30.414661 & 6.04 & 0.20 $\pm$ 0.03 & -19.07 $\pm$ 0.06\\
189 & 3.592944 & -30.413328 & 5.68 & 0.11 $\pm$ 0.04 & -16.44 $\pm$ 0.14\\
397 & 3.603215 & -30.410353 & 6.31 & 0.24 $\pm$ 0.04 & -17.08 $\pm$ 0.10\\
437 & 3.604567 & -30.409365 & 5.87 & 0.09 $\pm$ 0.04 & -15.56 $\pm$ 0.22\\
446 & 3.604751 & -30.409292 & 6.02 & 0.36 $\pm$ 0.07 & -16.61 $\pm$ 0.13\\
561 & 3.606384 & -30.407278 & 6.37 & 0.06 $\pm$ 0.03 & -17.85 $\pm$ 0.10\\
855 & 3.601075 & -30.403990 & 6.02 & 0.33 $\pm$ 0.13 & -17.48 $\pm$ 0.13\\
863 & 3.570069 & -30.403721 & 5.87 & 0.25 $\pm$ 0.04 & -18.29 $\pm$ 0.07\\
1622 & 3.586268 & -30.392712 & 5.86 & <0.05  & -15.15 $\pm$ 0.39\\
1989 & 3.606223 & -30.386644 & 5.59 & 0.35 $\pm$ 0.03 & -19.10 $\pm$ 0.04\\
2002 & 3.576895 & -30.386328 & 6.46 & 0.03 $\pm$ 0.02 & -15.99 $\pm$ 0.16\\
2007 & 3.588642 & -30.386244 & 5.82 & 0.07 $\pm$ 0.07 & -16.83 $\pm$ 0.29\\
2202 & 3.603425 & -30.383219 & 5.86 & 0.10 $\pm$ 0.07 & -17.34 $\pm$ 0.11\\
2434 & 3.590547 & -30.379766 & 5.82 & 0.39 $\pm$ 0.15 & -16.97 $\pm$ 0.16\\
2452 & 3.590790 & -30.379410 & 5.74 & 0.97 $\pm$ 0.10 & -18.34 $\pm$ 0.09\\
20214 & 3.572779 & -30.413527 & 5.84 & 0.50 $\pm$ 0.21 & -16.04 $\pm$ 0.23\\
20295 & 3.591288 & -30.412630 & 6.06 & 0.12 $\pm$ 0.10 & -15.54 $\pm$ 0.44\\
20785 & 3.605206 & -30.407106 & 5.85 & 0.19 $\pm$ 0.05 & -15.41 $\pm$ 0.30\\
21380 & 3.568028 & -30.401043 & 6.10 & 0.14 $\pm$ 0.10 & -18.06 $\pm$ 0.22\\
21582 & 3.589428 & -30.398497 & 5.88 & 0.04 $\pm$ 0.02 & -11.88 $\pm$ 0.34\\
21690 & 3.572050 & -30.397192 & 5.54 & <0.35  & -15.52 $\pm$ 0.40\\
22064 & 3.592817 & -30.393698 & 5.84 & <0.03  & -14.05 $\pm$ 0.32\\
22392 & 3.587827 & -30.390039 & 5.72 & <0.04  & -15.04 $\pm$ 0.22\\
22542 & 3.593637 & -30.388189 & 6.05 & 0.04 $\pm$ 0.04 & -15.34 $\pm$ 0.29\\
23153 & 3.597839 & -30.381126 & 6.11 & 0.25 $\pm$ 0.12 & -16.59 $\pm$ 0.26\\
23460 & 3.575475 & -30.408583 & 6.10 & 0.07 $\pm$ 0.06 & -16.14 $\pm$ 0.21\\

\hline
MACS J0416.1-2403\\
13 & 64.047569 & -24.097076 & 5.70 & 0.30 $\pm$ 0.08 & -20.85 $\pm$ 0.05\\
111 & 64.038246 & -24.092775 & 6.83 & 0.05 $\pm$ 0.02 & -19.35 $\pm$ 0.06\\
139 & 64.039261 & -24.091812 & 6.70 & 1.04 $\pm$ 0.10 & -19.86 $\pm$ 0.06\\
220 & 64.034561 & -24.089703 & 5.82 & 0.22 $\pm$ 0.09 & -17.75 $\pm$ 0.15\\
247 & 64.040497 & -24.088909 & 7.13 & 0.58 $\pm$ 0.14 & -18.95 $\pm$ 0.10\\
271 & 64.039551 & -24.088541 & 5.93 & 0.54 $\pm$ 0.05 & -19.58 $\pm$ 0.06\\
355 & 64.038498 & -24.087040 & 6.24 & 0.39 $\pm$ 0.15 & -17.93 $\pm$ 0.17\\
735 & 64.054390 & -24.081661 & 6.33 & 1.18 $\pm$ 0.09 & -21.31 $\pm$ 0.04\\
1827 & 64.050858 & -24.066542 & 5.91 & 0.16 $\pm$ 0.13 & -18.15 $\pm$ 0.14\\
1829 & 64.051079 & -24.066517 & 5.96 & <0.14  & -16.87 $\pm$ 0.20\\
2179 & 64.047852 & -24.062069 & 6.15 & 0.22 $\pm$ 0.04 & -18.12 $\pm$ 0.07\\
2204 & 64.039978 & -24.061810 & 6.30 & 0.41 $\pm$ 0.12 & -17.52 $\pm$ 0.13\\
2236 & 64.037483 & -24.061234 & 5.97 & 0.12 $\pm$ 0.06 & -17.37 $\pm$ 0.13\\
2240 & 64.047150 & -24.061132 & 5.75 & 0.24 $\pm$ 0.17 & -17.05 $\pm$ 0.07\\
2324 & 64.044479 & -24.059311 & 6.29 & <0.34  & -15.22 $\pm$ 0.25\\
2337 & 64.043571 & -24.059002 & 6.14 & <0.65  & -16.58 $\pm$ 0.16\\
2411 & 64.042946 & -24.057184 & 5.91 & <0.06  & -16.26 $\pm$ 0.20\\
20065 & 64.042404 & -24.095959 & 5.80 & <0.13  & -16.74 $\pm$ 0.35\\
20103 & 64.042519 & -24.094933 & 5.78 & 0.16 $\pm$ 0.12 & -16.64 $\pm$ 0.33\\
20470 & 64.037987 & -24.086891 & 5.87 & 0.18 $\pm$ 0.09 & -15.89 $\pm$ 0.49\\
20594 & 64.037933 & -24.085415 & 6.04 & 0.26 $\pm$ 0.15 & -15.98 $\pm$ 0.41\\
20618 & 64.040710 & -24.085100 & 5.91 & 0.27 $\pm$ 0.10 & -15.82 $\pm$ 0.44\\
20722 & 64.022423 & -24.083601 & 6.15 & 0.05 $\pm$ 0.02 & -14.32 $\pm$ 0.25\\
20914 & 64.037971 & -24.081451 & 5.87 & <0.05  & -14.22 $\pm$ 0.52\\
21216 & 64.020332 & -24.078485 & 5.75 & <0.19  & -15.81 $\pm$ 0.35\\
21650 & 64.046074 & -24.073437 & 6.10 & <0.33  & -16.51 $\pm$ 0.32\\
21659 & 64.019035 & -24.073301 & 6.36 & 2.61 $\pm$ 0.64 & -17.11 $\pm$ 0.34\\
21869 & 64.049286 & -24.070938 & 6.15 & 0.71 $\pm$ 0.16 & -17.35 $\pm$ 0.22\\
22453 & 64.050079 & -24.064102 & 5.79 & <0.04  & -15.64 $\pm$ 0.23\\
22971 & 64.037361 & -24.057997 & 5.99 & 0.23 $\pm$ 0.12 & -17.07 $\pm$ 0.23\\

\hline
\end{tabular}

\end{table*}

\begin{table*}
\caption{(Continued)}
\setcounter{table}{1}
\label{tab:fits-z67}
\begin{tabular}{cccccc}
\hline
ID & RA & Dec & z$_\text{photo} $ & r$_e$/kpc & M$_\text{UV}$ \\
\hline
 MACS J1149.5+2223\\
354 & 177.393036 & 22.381325 & 5.66 & <0.18  & -18.03 $\pm$ 0.15\\
404 & 177.388382 & 22.382391 & 6.20 & 0.24 $\pm$ 0.05 & -19.65 $\pm$ 0.08\\
433 & 177.384644 & 22.382761 & 6.16 & <0.16  & -18.56 $\pm$ 0.13\\
825 & 177.425262 & 22.386383 & 6.61 & 0.49 $\pm$ 0.01 & -23.14 $\pm$ 0.01\\
1243 & 177.384537 & 22.392721 & 5.92 & 1.06 $\pm$ 0.15 & -18.72 $\pm$ 0.11\\
1494 & 177.382538 & 22.395403 & 5.76 & 0.87 $\pm$ 0.05 & -19.89 $\pm$ 0.05\\
2364 & 177.387711 & 22.405834 & 5.71 & 4.17 $\pm$ 3.14 & -16.35 $\pm$ 0.23\\
2368 & 177.418457 & 22.405724 & 5.94 & 0.47 $\pm$ 0.14 & -19.52 $\pm$ 0.07\\
2410 & 177.418625 & 22.406404 & 6.00 & 0.63 $\pm$ 0.09 & -19.02 $\pm$ 0.10\\
2535 & 177.409424 & 22.408207 & 5.54 & 0.29 $\pm$ 0.05 & -19.03 $\pm$ 0.07\\
2619 & 177.403015 & 22.409502 & 5.79 & 0.32 $\pm$ 0.06 & -18.71 $\pm$ 0.07\\
2747 & 177.407974 & 22.411522 & 6.17 & < 0.35 & -17.92 $\pm$ 0.23\\
2966 & 177.412933 & 22.413643 & 6.40 & 0.07 $\pm$ 0.04 & -18.64 $\pm$ 0.11\\
3027 & 177.402725 & 22.414614 & 5.83 & 0.18 $\pm$ 0.09 & -17.74 $\pm$ 0.16\\
3054 & 177.412018 & 22.415779 & 5.79 & 1.03 $\pm$ 0.02 & -21.14 $\pm$ 0.02\\
3162 & 177.414688 & 22.417059 & 6.07 & 0.55 $\pm$ 0.11 & -18.51 $\pm$ 0.14\\
20570 & 177.393036 & 22.381321 & 5.75 & 0.18 $\pm$ 0.13 & -18.00 $\pm$ 0.13\\
21231 & 177.397278 & 22.387806 & 5.85 & <0.98  & -16.34 $\pm$ 0.37\\
22626 & 177.386780 & 22.401293 & 5.96 & <0.55 & -17.39 $\pm$ 0.17\\
22717 & 177.393097 & 22.402140 & 5.80 & 0.16 $\pm$ 0.11 & -17.41 $\pm$ 0.14\\
23096 & 177.380600 & 22.405464 & 5.64 & <0.27  & -17.48 $\pm$ 0.19\\
23485 & 177.403015 & 22.409500 & 5.91 & 0.23 $\pm$ 0.05 & -18.76 $\pm$ 0.07\\
23874 & 177.412933 & 22.413645 & 6.46 & 0.05 $\pm$ 0.04 & -18.69 $\pm$ 0.08\\
24039 & 177.412018 & 22.415777 & 5.98 & 1.01 $\pm$ 0.02 & -21.20 $\pm$ 0.02\\
\hline

MACS J0717.5+3745\\
82 & 109.389275 & 37.724854 & 6.05 & 0.56 $\pm$ 0.05 & -18.59 $\pm$ 0.08\\
136 & 109.381462 & 37.726364 & 5.58 & 0.90 $\pm$ 0.08 & -19.90 $\pm$ 0.05\\
203 & 109.379547 & 37.728397 & 6.25 & 2.22 $\pm$ 0.24 & -20.30 $\pm$ 0.08\\
208 & 109.378731 & 37.728226 & 6.17 & 4.53 $\pm$ 1.47 & -19.03 $\pm$ 0.21\\
246 & 109.378822 & 37.729141 & 5.60 & 0.89 $\pm$ 0.10 & -19.57 $\pm$ 0.11\\
250 & 109.378723 & 37.729404 & 7.20 & 1.39 $\pm$ 0.69 & -18.25 $\pm$ 0.28\\
332 & 109.376572 & 37.731480 & 6.31 & 1.27 $\pm$ 0.22 & -19.09 $\pm$ 0.15\\
336 & 109.381050 & 37.731613 & 6.10 & <0.10 & -18.21 $\pm$ 0.10\\
356 & 109.378174 & 37.731903 & 6.16 & 0.50 $\pm$ 0.21 & -16.93 $\pm$ 0.23\\
374 & 109.377205 & 37.732182 & 6.25 & 0.20 $\pm$ 0.07 & -18.04 $\pm$ 0.17\\
392 & 109.376816 & 37.732525 & 6.03 & 0.75 $\pm$ 0.18 & -17.91 $\pm$ 0.24\\
471 & 109.412857 & 37.733803 & 6.53 & 0.05 $\pm$ 0.01 & -15.10 $\pm$ 0.08\\
510 & 109.413666 & 37.734642 & 6.46 & <0.01 & -13.87 $\pm$ 0.08\\
640 & 109.376984 & 37.736454 & 6.02 & 0.58 $\pm$ 0.07 & -18.51 $\pm$ 0.09\\
774 & 109.379036 & 37.738388 & 5.93 & 0.18 $\pm$ 0.09 & -16.59 $\pm$ 0.16\\
797 & 109.391907 & 37.737995 & 5.97 & 2.49 $\pm$ 1.08 & -17.97 $\pm$ 0.04\\
813 & 109.418655 & 37.738789 & 5.90 & <0.07 & -16.62 $\pm$ 0.18\\
1095 & 109.377609 & 37.741798 & 5.73 & 0.19 $\pm$ 0.06 & -17.12 $\pm$ 0.11\\
1202 & 109.390739 & 37.742214 & 6.13 & 0.88 $\pm$ 0.04 & -20.19 $\pm$ 0.03\\
1737 & 109.408783 & 37.748375 & 6.04 & 1.05 $\pm$ 0.43 & -14.98 $\pm$ 0.13\\
1841 & 109.380402 & 37.749458 & 5.74 & 0.26 $\pm$ 0.07 & -17.76 $\pm$ 0.13\\
2090 & 109.386215 & 37.751919 & 6.31 & 1.03 $\pm$ 0.09 & -19.47 $\pm$ 0.06\\
2169 & 109.384567 & 37.753006 & 5.86 & 0.49 $\pm$ 0.04 & -21.14 $\pm$ 0.02\\
2312 & 109.409065 & 37.754684 & 6.12 & 0.11 $\pm$ 0.02 & -18.51 $\pm$ 0.05\\
2368 & 109.412941 & 37.755592 & 5.75 & 1.15 $\pm$ 0.17 & -18.85 $\pm$ 0.16\\
2883 & 109.399109 & 37.764957 & 6.37 & 0.74 $\pm$ 0.08 & -19.39 $\pm$ 0.06\\
20950 & 109.370537 & 37.738941 & 5.59 & 1.20 $\pm$ 0.85 & -16.23 $\pm$ 0.26\\
23199 & 109.403320 & 37.765663 & 5.88 & 0.14 $\pm$ 0.18 & -16.73 $\pm$ 0.23\\
23588 & 109.385132 & 37.758907 & 6.08 & 0.04 $\pm$ 0.02 & -15.20 $\pm$ 0.29\\

\hline

\end{tabular}

\end{table*}

\begin{table*}
\caption{(Continued)}
\setcounter{table}{2}
\label{tab:fits-z67}
\begin{tabular}{cccccc}
\hline
ID & RA & Dec & z$_\text{photo}$ & r$_e$/kpc & M$_\text{UV}$ \\
\hline
Abell S1063\\
165 & 342.201924 & -44.549906 & 5.93 & 2.53 $\pm$ 0.73 & -19.29 $\pm$ 0.08\\
242 & 342.194600 & -44.548404 & 5.83 & 3.40 $\pm$ 3.27 & -18.55 $\pm$ 0.17\\
247 & 342.177057 & -44.548331 & 6.16 & 1.30 $\pm$ 0.19 & -19.55 $\pm$ 0.09\\
325 & 342.177695 & -44.546933 & 6.28 & 0.90 $\pm$ 0.11 & -19.59 $\pm$ 0.05\\
438 & 342.173075 & -44.545251 & 6.29 & 0.53 $\pm$ 0.29 & -18.18 $\pm$ 0.13\\
617 & 342.193776 & -44.542721 & 6.44 & <0.37  & -17.69 $\pm$ 0.19\\
626 & 342.176131 & -44.542668 & 6.21 & 0.23 $\pm$ 0.05 & -17.92 $\pm$ 0.08\\
652 & 342.167993 & -44.542512 & 6.04 & 0.07 $\pm$ 0.02 & -17.28 $\pm$ 0.04\\
1033 & 342.190892 & -44.537462 & 6.11 & 0.23 $\pm$ 0.01 & -20.42 $\pm$ 0.02\\
1142 & 342.198542 & -44.535311 & 5.52 & <0.50 & -16.51 $\pm$ 0.22\\
1143 & 342.183924 & -44.535317 & 6.11 & 0.20 $\pm$ 0.10 & -17.13 $\pm$ 0.21\\
1233 & 342.181055 & -44.534609 & 6.11 & 0.46 $\pm$ 0.02 & -21.29 $\pm$ 0.03\\
1690 & 342.164364 & -44.530232 & 6.12 & 0.65 $\pm$ 0.16 & -18.00 $\pm$ 0.11\\
1717 & 342.189053 & -44.530028 & 6.11 & 0.77 $\pm$ 0.02 & -21.64 $\pm$ 0.03\\
2073 & 342.173329 & -44.525929 & 5.91 & 0.07 $\pm$ 0.03 & -17.34 $\pm$ 0.07\\
2592 & 342.171298 & -44.519801 & 6.11 & 0.32 $\pm$ 0.41 & -20.09 $\pm$ 0.03\\
2608 & 342.172002 & -44.519331 & 6.55 & 0.55 $\pm$ 0.39 & -17.82 $\pm$ 0.17\\
2932 & 342.169900 & -44.515155 & 6.74 & <1.35  & -17.73 $\pm$ 0.24\\
2955 & 342.165543 & -44.514668 & 6.14 & 1.61 $\pm$ 1.04 & -17.68 $\pm$ 0.17\\
3001 & 342.173965 & -44.514089 & 5.98 & 1.00 $\pm$ 0.43 & -17.98 $\pm$ 0.18\\

\hline
 Abell 370\\
729 & 39.968278 & -1.597248 & 6.80 & <0.62  & -17.82 $\pm$ 0.38\\
2026 & 39.985087 & -1.578782 & 6.75 & 1.97 $\pm$ 0.56 & -18.53 $\pm$ 0.40\\
2356 & 39.986285 & -1.575199 & 5.62 & 0.12 $\pm$ 0.09 & -18.23 $\pm$ 0.21\\
2843 & 39.985811 & -1.571304 & 6.32 & 0.45 $\pm$ 0.10 & -18.92 $\pm$ 0.16\\
3238 & 39.982590 & -1.566664 & 6.65 & <0.09  & -16.99 $\pm$ 0.27\\
3241 & 39.980269 & -1.566761 & 6.17 & 0.09 $\pm$ 0.04 & -18.86 $\pm$ 0.09\\
3252 & 39.980109 & -1.566731 & 6.10 & 0.60 $\pm$ 0.08 & -19.12 $\pm$ 0.10\\
3822 & 39.978189 & -1.558981 & 6.77 & 0.24 $\pm$ 0.03 & -20.26 $\pm$ 0.04\\
3825 & 39.978067 & -1.558955 & 6.77 & 0.23 $\pm$ 0.02 & -20.18 $\pm$ 0.04\\

\hline
\end{tabular}

\end{table*}

\begin{table*}
\caption{Same as Table~\ref{tab:fits-z67}, for galaxies at $z\sim8.5$.}
\label{tab:fits-z8}
\begin{tabular}{cccccc}
\hline\hline
ID & RA & Dec & z$_\text{photo} $ & r$_e$/kpc & M$_\text{UV}$ \\
\hline

Abell 2744\\
409 & 3.592160 & -30.409914 & 7.66 & <0.16 & -16.56 $\pm$ 0.19\\
548 & 3.576798 & -30.407442 & 8.56 & 0.31 $\pm$ 0.13 & -17.77 $\pm$ 0.20\\
2036 & 3.596087 & -30.385836 & 8.32 & 0.36 $\pm$ 0.04 & -19.37 $\pm$ 0.07\\
2248 & 3.603863 & -30.382261 & 8.02 & 0.50 $\pm$ 0.07 & -20.04 $\pm$ 0.07\\
2257 & 3.598123 & -30.382393 & 7.53 & 0.34 $\pm$ 0.15 & -17.74 $\pm$ 0.18\\
2261 & 3.603996 & -30.382309 & 7.97 & 1.07 $\pm$ 0.18 & -19.32 $\pm$ 0.10\\
2264 & 3.603382 & -30.382252 & 8.29 & 0.94 $\pm$ 0.04 & -20.29 $\pm$ 0.06\\
2316 & 3.605061 & -30.381470 & 7.66 & 0.52 $\pm$ 0.18 & -18.60 $\pm$ 0.19\\
2346 & 3.606460 & -30.380995 & 7.79 & 0.45 $\pm$ 0.08 & -19.86 $\pm$ 0.06\\
2365 & 3.604516 & -30.380468 & 8.20 & 0.35 $\pm$ 0.03 & -20.84 $\pm$ 0.05\\
20236 & 3.572523 & -30.413267 & 8.46 & 1.03 $\pm$ 0.21 & -18.01 $\pm$ 0.24\\
21573 & 3.579663 & -30.398682 & 7.75 & 0.05 $\pm$ 0.03 & -15.57 $\pm$ 0.19\\
22347 & 3.588001 & -30.390648 & 8.80 & 0.12 $\pm$ 0.07 & -16.62 $\pm$ 0.20\\
\hline
MACS J0416.1-2403\\
99 & 64.039162 & -24.093182 & 8.66 & 0.76 $\pm$ 0.04 & -20.54 $\pm$ 0.05\\
218 & 64.027420 & -24.089979 & 7.68 & 1.26 $\pm$ 0.04 & -19.83 $\pm$ 0.03\\
286 & 64.037567 & -24.088116 & 8.14 & 0.24 $\pm$ 0.10 & -18.22 $\pm$ 0.17\\
726 & 64.047981 & -24.081671 & 8.32 & 0.29 $\pm$ 0.03 & -19.94 $\pm$ 0.06\\
743 & 64.048058 & -24.081427 & 8.94 & 0.61 $\pm$ 0.07 & -20.24 $\pm$ 0.06\\
1253 & 64.019333 & -24.075153 & 7.66 & 2.65 $\pm$ 0.50 & -18.79 $\pm$ 0.22\\
1997 & 64.049583 & -24.064596 & 8.10 & 0.15 $\pm$ 0.05 & -17.82 $\pm$ 0.17\\
20465 & 64.049088 & -24.086967 & 7.53 & 2.54 $\pm$ 0.73 & -18.96 $\pm$ 0.22\\
21169 & 64.041405 & -24.078926 & 8.03 & <0.11 & -17.11 $\pm$ 0.23\\
21864 & 64.024414 & -24.070963 & 7.64 & 1.12 $\pm$ 0.44 & -18.52 $\pm$ 0.21\\
21997 & 64.039139 & -24.069435 & 7.95 & 0.25 $\pm$ 0.06 & -15.71 $\pm$ 0.23\\
22035 & 64.038971 & -24.069519 & 7.65 & 0.62 $\pm$ 0.08 & -18.56 $\pm$ 0.14\\
22746 & 64.046509 & -24.061630 & 8.36 & 0.40 $\pm$ 0.20 & -17.27 $\pm$ 0.23\\
\hline
MACS J1149.5+2223\\
2316 & 177.407684 & 22.405251 & 7.93 & <0.22  & -16.91 $\pm$ 0.27\\
22863 & 177.388184 & 22.403442 & 7.66 & <0.70 & -14.83 $\pm$ 0.43\\
\hline
MACS J0717.5+3745\\
23536 & 109.400124 & 37.746799 & 8.45 & 0.51 $\pm$ 0.67 & -17.90 $\pm$ 0.06\\
\hline
Abell S1063\\
2048 & 342.182253 & -44.526073 & 8.81 & 0.10 $\pm$ 0.02 & -16.40 $\pm$ 0.13\\
2086 & 342.179829 & -44.525664 & 8.85 & 0.13 $\pm$ 0.02 & -16.67 $\pm$ 0.09\\
\hline
Abell 370\\
849 & 39.966126 & -1.594771 & 7.70 & 0.07 $\pm$ 0.03 & -16.68 $\pm$ 0.19\\
2867 & 39.992152 & -1.570892 & 8.20 & 0.57 $\pm$ 0.13 & -19.51 $\pm$ 0.15\\
\hline
\end{tabular}

\end{table*}

\section*{Acknowledgements}
The research leading to these results has received funding from the European Union Seventh Framework Programme (FP7/2007-2013) under grant agreement No. 312725.
This work utilizes gravitational lensing models produced by PIs Brada\v{c}, Natarajan \& Kneib (CATS), Merten \& Zitrin, Sharon, Williams groups. This lens modeling was partially funded by the HST Frontier Fields program conducted by STScI. STScI is operated by the Association of Universities for Research in Astronomy, Inc. under NASA contract NAS 5-26555. The lens models were obtained from the Mikulski Archive for Space Telescopes (MAST).  
LY acknowledges support by JSPS KAKENHI Grant Number JP 21F21325.
LY and TT acknowledge support by NASA through grant JWST-ERS-1324.
MB acknowledges support provided by the National Science Foundation under Grant No. AST-1815458. 
The authors thank Karl Glazebrook and Danilo Marchesini for several discussions that helped shaped the manuscript.

\section*{Data Availability} 

The SEDs, cutouts, and all ancillary information of galaxies used in this article can be found on the ASTRODEEP CDS Interface at \url{http://www.astrodeep.eu/cds-interface/}.



\bibliographystyle{mnras}
\bibliography{reference} 




\bsp	
\label{lastpage}
\end{document}